\documentclass[11pt,twoside]{article}
\usepackage{asp2010}

\resetcounters

\begin{document}

\title{The kinematics and morphology of PNe with close binary nuclei}
\author{L\'{o}pez, J. A.$^1$, Garc\'{\i}a-D\'{\i}az, Ma. T.$^1$, Richer, M. G$^1$,  Lloyd, M.$^2$ and  Meaburn, J.$^2$
\affil{$^1$Instituto de Astronom\'{\i}a, Universidad Nacional Aut\'onoma de M\'exico, Apdo. Postal 877, Ensenada, Baja California, 22800, M\'exico}
\affil{$^2$Jodrell Bank Centre for Astrophysics, University of Manchester, Oxford Road, Manchester M 13 9PL, UK}}

\begin{abstract}
We have obtained images and long-slit, spatially resolved echelle spectra for twenty four planetary nebulae (PNe) that have confirmed close binary nuclei. The sample shows a variety of morphologies, however toroids or dense equatorial density enhancements are identified, both in the imagery and the spectra, as the common structural component. These toroids are thought to be the remnant fingerprints of the post common envelope phase. Based on the characteristics of the present sample we suggest a list of additional PNe that are likely to host close binary nuclei.
\end{abstract}

\section*{Introduction}
The presence of fast, collimated, bipolar outflows or jets, and point-symmetric and poly-polar structures in PNe has led for nearly two decades to the suggestion that close binary nuclei should be playing a decisive role in their formation and shaping (e. g. Soker \& Livio 1994). Furthermore, evolution paths and population synthesis calculations critically depend on the actual ratio of binary to single core PNe. However, finding these binary systems has proven to be an observational challenge, but after years of dedicated efforts a list of  PNe with reasonably confirmed close binary nuclei has been gradually obtained and this has been nicely summarized by De Marco (2009) and references therein. With this group of identified targets available we set out to investigate its detailed morphological and kinematical characteristics.

\section*{Observations}

High-resolution spectroscopic observations and monochromatic images were obtained at the Observatorio Astron\'omico Nacional at San Pedro M\'{a}rtir (SPM), Baja California,
M\'exico and at the Anglo - Australian Observatory. The observations were obtained with the Manchester Echelle Spectrometer (MES-SPM) on the SPM telescope and with UCLES  on the AAT for the southernmost targets.  In both cases we used a 90 \AA~ bandwidth filter to isolate the 87$^{th}$ order containing the H$\alpha$ and [N II] $\lambda$$\lambda ~$6548, 6584 \AA, nebular emission lines. The SPM spectra have a projected slit length of 5\farcm32 on the sky and those from UCLES approximately one arcmin. In both cases the spectral resolution is $\sim$ 11 km s$^{-1}$. For the SPM targets we also used MES in its imaging mode to  obtain monochromatic images. An additional source of high quality images has been provided by the web based planetary nebulae image catalogue (PNIC) compiled by Bruce Balick (www.astro.washington.edu/users/balick/PNIC/) and its modified version by Brent Miszalski (www.aao.gov.au/local/www/brent/PNIC/).

\section*{The sample}

The twenty four PNe observed in our sample are listed below, grouped by their main large-scale morphology, those marked with an asterisk contain collimated bipolar outflows.
\begin{itemize}
\item Ring, Toroid, Cylindrical : NGC 6337*, Hf 2-2, M 3-16, NGC 6026, A 63*, A 46, A 65, Sp 1, SuWt 2, NGC 2438, Ha Tr 4.

\item Elliptical with marked equatorial density enhancements: A 41, H 2-29, M 2-19, H 1-33.

\item Bipolar: PN G 136, NGC 2346, K 1-2*.

\item Irregular, Diffuse, Double Shell, Filamentary Bubble: Ds 1, Lo Tr 5, Lo Tr 1, NGC 1514, HFG 1, A 35.
\end{itemize}
The first group comprises 46\% of the sample, whereas the second group 17\%, the third 12\% and the fourth 25\%. It is clear that this is a mixed sample and prone to the uncertainties of small numbers statistics. Yet this sample seems strongly dominated by rings, toroids, cylindrical shapes and pronounced equatorial density enhancements, somehow in contrast with the expectations for this particular group as described in the introduction. These results agree with those from Frew and Parker (2007) derived from a smaller sample. Long-slit spectra serve as an excellent tool to unveil the presence of collimated outflows and also toroidal or ring-type structures from the additional information  provided in the emission line profiles (see Figure 1). The kinematics of these PNe does not show particularly high expansion velocities, they all seem rather standard, with expansion velocities of the order of $V_{exp} = 30 \pm 5$ km s$^{-1}$. Only three members of this sample are known to host bipolar, collimated outflows, NGC 6337, A 63 and K 1-2, being NGC 6337 the most complex one  (Garc\'{\i}a-D\'{\i}az et al. 2009).  Other examples of this group have been discussed during this meeting, such as Hen 2-428 (Santander-Garc\'{\i}a et al. 2010) and the necklace nebula, (R. Corradi et al. (2010), both are dominated by toroidal structures at the equator, in the latter a collimated bipolar outflow has also been found.   

\begin{figure}[!ht]
\plotone{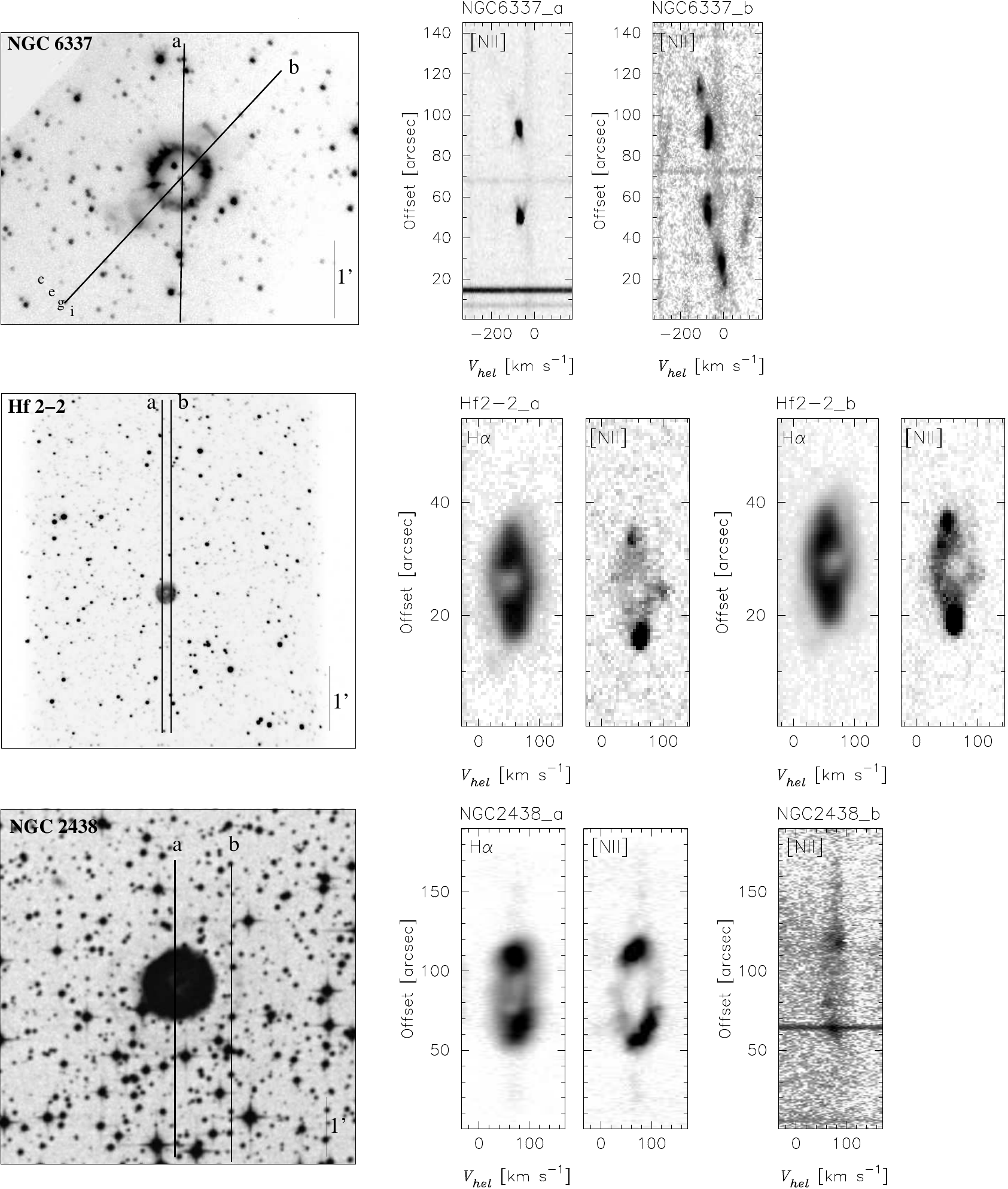}
\caption{Examples of images and long-slit spectra for three PNe with close binary nuclei. The location of the slits is indicated on the image for each object.  NGC 6337 is shown in the top panel. This nebula is dominated by a ring structure which is clearly replicated in the spectra. This object also has a complex collimated outflow. In the middle panel is Hf 2-2 that also shows clearly the toroidal structure in the image and the spectra from both slits. The bottom panel shows NGC 2348 deliberately in an image from the digital sky survey that only shows a filled envelope, this case exemplifies how the ring structure is clearly revealed again in the long-slit spectra. }
\end{figure}

\section{Discussion}
From a similar sample, Miszalski et al. (2009) consider that PNe with close binary nuclei show a clear tendency for bipolar shapes once inclination effects are taken into account. 
The present sample size does not allow yet to draw firm conclusions on the influence of close binary nuclei on the shaping and evolution of PNe. New members of this group are currently being actively sought with modern techniques and it is anticipated that many more members of this group will be found in the near future. From the diversity of the members studied in the present sample it can only be expected that more PNe within the categories described above will accumulate until a clearer trend coupled to a better understanding  of the binary interaction processes on the shaping mechanisms of the PN can be envisaged.

For the moment, the simplest interpretation to the present results is to assume that PNe with close binary nuclei are post common envelope systems that have predominantly ejected the primary's envelope along the equatorial plane, producing the equatorial density enhancement or toroidal structures observed. Gravitational focusing along the equator from a close low-mass  companion can also play an important role, (see e.g. the review by Podsialowski (2007) and Nordhaus, Blackman \& Frank (2006)). Once the toroid is formed, at the early stages of PN evolution, the fast isotropic stellar wind is deviated towards the polar regions. Depending on the amount of envelope matter that has been deposited on the equator (i.e. the efficiency of the envelope ejection) the resultant main morphology will appear as toroid-dominated, elliptical with a thick waist or bipolar, with the first two gradually evolving towards the latter as a function of time. It is important to note that under this interpretation the common envelope process is mainly responsible for distributing the isotropic envelope over the equator producing the toroidal structures, not  an axysimmetric (e.g. bipolar) nebula {\it per se}. This large-scale mechanism does not consider the production of fast, bipolar, collimated outflows or point-symmetric and poly-polar structures that must be formed through additional processes, that are  also likely related to detailed conditions of the binary nature, its mechanism of interaction and the presence of collimating magnetic fields. Thus for a better evaluation and prediction of the likely effects of binary nuclei on PNe shaping we need a larger sample of confirmed close binary nuclei. Additionally, we need a much better understanding of the detailed outcome of the common envelope process as a function of mass ratio and its dependency on the moment at which this happens during the system's evolution. We also need a better understanding of the shaping effects on the nebula by wider binaries, e.g. Frankowski \& Jorissen (20007). For the time being, if the characteristics of the members in the present sample are indeed indicative of the likely presence of close binary nuclei, we suggest to study the following candidates:  K 1-9, A 13, Lo 18, SuWt 3, A 47, He 2-70, Hf 4, He 2-163 and Al 1.

\acknowledgements
We gratefully acknowledge the financial support from DGAPA-UNAM through grant IN116908


\end{document}